\documentclass[reprint,superscriptaddress,onecolumn, aps, prl]{revtex4-1}

\usepackage{graphicx,psfrag}
\usepackage{epstopdf}
\usepackage{amsmath} 
\usepackage{bm}
\usepackage{amssymb}
\usepackage{subfigure}
\usepackage{color}
\usepackage[pagewise]{lineno}

\begin{document}

\title{\large \bf Double-zero-index structural waveguides}

\author{Hongfei Zhu}
\email{Hongfei.Zhu.44@nd.edu}
\affiliation{Department of Aerospace and Mechanical Engineering, University of Notre Dame, Notre Dame, IN 46556, USA}
\author {Fabio Semperlotti}
\email{To whom correspondence should be addressed: fsemperl@purdue.edu}
\affiliation{Ray W. Herrick Laboratories, School of Mechanical Engineering, Purdue University, West Lafayette, Indiana 47907, USA}

\date{\today}

\begin{abstract}
	
We report on the theoretical and experimental realization of a double-zero-index elastic waveguide and on the corresponding acoustic cloacking and supercoupling effects. The proposed waveguide uses geometric tapers in order to induce Dirac-like cones at $\vec{k}=0$ due to accidental degeneracy. The nature of the degeneracy is explored by a $k\cdot p$ perturbation method adapted to thin structural waveguides. Results confirm the linear nature of the dispersion around the degeneracy and the possibility to map the material to effective medium properties. Effective parameters, numerically extracted using boundary medium theory, confirm that the phononic waveguide maps into a double zero index material. Numerical and experimental results confirm the expected cloacking and supecoupling effects typical of zero index materials.

\end{abstract}

\maketitle

\section{Introduction}

The concept of acoustic metamaterials \cite{cui} has rapidly emerged as a powerful alternative to design materials and structures exhibiting unexpected dynamic properties typically not achievable in natural materials. In less than two decades, this concept has allowed a drastic expansion of the materials design space enabling novel applications involving acoustic wave management and control. Properties such as acoustic bandgaps  \cite{Liu,Yang,Helios,Garcia,Vasseur}, focusing \cite{focusing1,focusing2,focusing3,focusing4,focusing5,focusing6,focusing7,Zhu}, collimation \cite{collimation1,collimation2,collimation3,collimation4}, sub-wavelength resolution \cite{sub1,sub2,sub3,sub4,Semperlotti}, and negative refraction \cite{Morvan,Pierre}, have been discovered and studied in depth. More recently, researchers have shown a rather new and exciting property of these materials consisting in their ability to achieve near-zero effective parameters. This class of materials was first formulated for electromagnetic waves where epsilon-near-zero (ENZ), mu-near-zero (MNZ), and epsilon-and-mu-near-zero (EMNZ) properties were first obtained. Applications included antenna designs with high directivity \cite{Alu,Stefan} and enhanced radiation efficiency \cite{Alu2,Soric}, as well as the realization of unconventional tunneling of electromagnetic energy within ultra-thin subwavelength channels or bends \cite{Ziolkowski,Silver,Silver2}. Among the most peculiar characteristics of these materials, we mention the independence of the phase from the propagation distance. This means that a wave entering a double-zero material emerges on the other side having the exact same phase as the input. In addition, double-zero materials are also characterized by a high level of transmissibility, ideally acting as a non-reflective waveguide even in presence of sharp impedance discontinuities.

While materials with near-zero permittivity are available in nature (e.g. some noble metals, doped semiconductors \cite{adams}, polar dielectrics \cite{Silver1}, transparent conducting oxide (TCOs) \cite{Naik}), in acoustics near-zero density and elastic compliance must be achieved as effective quantities by leveraging the local dynamic response of the medium. In the past few years, some acoustic metamaterials were reported to exhibit single zero effective parameters, such as near zero density \cite{DNZ1,DNZ2,DNZ3}. We note that this behavior is the exact counterpart to single zero electromagnatic materials, such as the ENZ. Although these materials offered good control on the phase, they suffered from low-transmissibility due to an intrinsic impedance mismatch between the host and the zero effective density medium. Electromagnetic double zero materials were designed specifically to alleviate this limitation on the transmission properties. However, designing acoustic media with double zero effective properties is not a trivial task given that, as previously mentioned, they are not readily available in nature.

Recent studies on photonic and phononic crystals \cite{Dirac1,Dirac2,Dirac3,Dirac4,Dirac5,Dirac6,Dirac7,Dirac8,Dirac9} revealed that when a Dirac-like Cone (DC) can be obtained at the center of the Brillouin zone, such lattice can be mapped into a double-zero refractive index material. This observation has drastically extended the possibility to design materials having near-zero effective properties. Nevertheless, while different applications of this basic concept were explored in photonics and phononics, there has been very little research targeting the implementation of these effective material properties in solids \cite{ZIMSOLID1,Dirac1}. The research on elastic phononic waveguides has been lagging behind even more and it currently counts no attempt of designing zero-index properties. In addition, the experimental implementation and validation of zero-index elastic media has not been reported in the scientific literature mostly due to the complexities associated with their design and fabrication.
 
In the present study, we report on the theoretical, numerical, and experimental realization of a structural phononic waveguide exhibiting double-zero-index-material (DZIM) behavior and capable of achieving acoustic cloaking and supercoupling.

The proposed design builds upon a class of metamaterials recently introduced by the authors \cite{ZhuPRB,ZhuPRL,ZhuJAP} and based on geometric tapers realized in a single-material system. The specific design employed in this study can be thought as an equivalent locally-resonant unit where an internal resonating core is embedded within a more compliant medium (i.e. the taper).

Geometrically tapered metamaterials \cite{ZhuPRB} exhibit Dirac-like Cones (DC) at the center of the Brillouin zone ($\Gamma$ point) that are the result of accidental degeneracies \cite{Meikp}. In other terms, the degeneracy is induced by the specific combination of the geometric parameters of the tapers and it is not protected by the underlying lattice structure (like, as an example, in graphene). The bands emanating from the three-fold degenerate point (the Dirac-like point) exhibit isotropic linear dispersion. We will show that these properties are the foundation that allows achieving double-zero effective properties in this class of materials. In particular, we will show that in the neighborhood of this degenerate point our waveguide exhibits simultaneous zero mass density and and zero reciprocal shear modulus (or, equivalently, infinite shear modulus).

\section{Results}
\subsection{Double-zero index waveguide via geometric tailoring}

\begin{figure}[h]
	\includegraphics[scale=0.42]{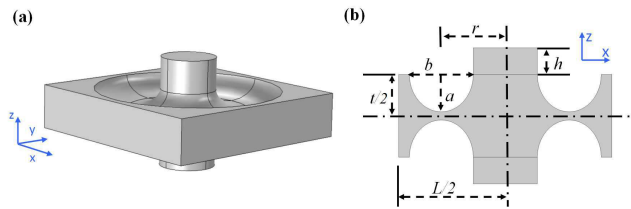}
	\caption{Schematic of (a) the fundamental tapered unit cell, and (b) its $x-z$ cross-section showing the main geometric parameters. The tapered geometry is symmetric with respect to the mid-plane of the plate.} \label{Fig1}
\end{figure}

The proposed phononic waveguide employs a tapered unit cell in a square lattice configuration ($C_{4v}$ symmetry). The unit cell consists of a square plate having an embedded elliptic torus-like taper and a (resonating) center mass. The unit is also symmetric with respect to the plate mid-plane (see Fig.\ref{Fig1}). The $x-z$ cross-section of the unit is shown in Fig.\ref{Fig1}b and shows the main geometric parameters where $t$ is the thickness of the element, $L$ is the lattice constant, $a$ and $b$ are the lengths of the minor and major axes of the ellipse, $r$ is the radius of the torus, and $h$ is the thickness of center mass. The unit was made out of aluminum with mass density $\rho=2700$ $kg/m^3$, Young's modulus $E=70$ GPa, and the Poisson's ratio $\nu=0.33$. 

\begin{figure}
	\includegraphics[scale=0.42]{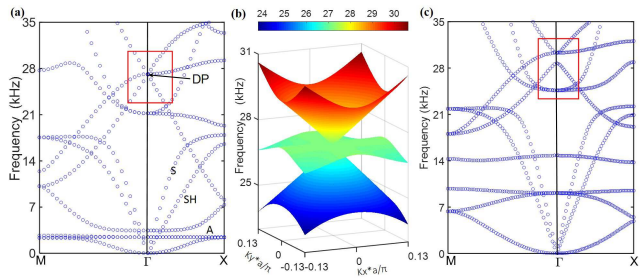}
	\caption{(a) The dispersion relations around $f =27.02$ kHz showing the existence of a triple-degenerate point and a Dirac-like cone (red box) for a given selection of the taper parameters. (b) Equi-Frequency-Surface plot corresponding to the frequency range around the DP and showing the formation of the Dirac-like cones. (c) When the geometric configuration is slightly perturbed (taper coefficient $a$ changed from 0.0039 to 0.0035) the Dirac-like cone opens up, indicating that the formation of the linear dispersions is due to an accidental degeneracy.} \label{Fig2}
\end{figure}

The dispersion relations for the proposed system were calculated using a commercial finite element solver (Comsol Multiphysics). Given the finite dimension of the unit cell in the thickness direction the dispersion curves are composed by symmetric (S), anti-symmetric (A), and shear horizontal (SH) guided Lamb modes. By using a proper selection of the geometric parameters (specifically $L=0.04$m, $t=0.008$m, $ a=0.0039$m, $b=0.007442$m, $r=0.012$m and $h=0.013$m), the band structure of the waveguide was tuned to exhibit a three-fold degenerate point (the Dirac-like Point, DP) at $f=27.04$ kHz and $\vec{k}=0$ (Fig.\ref{Fig2}a). Note that the branches emanating from the degenerate point are isotropic and linear and form two cones touching on their vertices at the DP (Fig. \ref{Fig2}b). The cones are made of $A_0$ modes having non-zero but constant group velocity and are intersected by an $A_0$ flat band at the DP.

The Dirac-like cone is the result of an accidental degeneracy, which can be confirmed by slightly perturbing the geometric parameters. By perturbing the torus-like taper $a$ (from 0.0039 to 0.0035) the cones separate (see Fig.\ref{Fig2}c) and the triple degenerate point splits into a non-degenerate and a doubly degenerate band. The corresponding eigenstates of the three degenerate modes are provided in Fig.\ref{Fig3}a which shows, from top to bottom, the lower cone, the flat band, and the upper cone.

\subsection{Analysis of the Dirac-like cones}

To further understand the origin of the Dirac-like dispersion, we extended the $\vec{k}\cdot\vec{p}$ method (well-known in electronic applications) to analyze our phononic system. This method was recently adopted by Mei\cite{Meikp} to analyze the Dirac and Dirac-like cones in two-dimensional phononic and photonic crystals. 

Our system is described by the Navier's equations with traction-free boundary conditions on the top and bottom surfaces of the waveguide. Imposing such boundary conditions when in presence of tapers creates non-trivial complexities due to the changing direction of the unit vector normal to the tapered surface. In order to avoid this issue, we resorted to an approach previously used to extend the three-dimensional plane wave expansion method to 2D phononic waveguides \cite{3DPWE,ZhuPRB}. The details of the method are provided in Supplementary. According to this method, the 2D waveguide is part of a layered three-dimensional periodic system that is constructed by alternating the waveguide and vacuum layers along the thickness direction. The resulting three-dimensional periodic medium can then be modeled as a layered bulk material according to the Navier's equations. The solution of the homogeneous problem provides the dispersions for the medium. Note that the use of the vacuum layer ensures that the periodic images of the waveguide along the thickness direction are dynamically decoupled from each other, therefore simply resulting in repeated roots in the dispersion calculation.

We can then write the general form of the Navier's equations for an inhomogeneous bulk medium as,
\begin{align}
\begin{split}
-\rho \omega^2\vec{U}=(\lambda+\mu)\nabla(\nabla\cdot\vec{U})+\mu\nabla^2\vec{U}+\nabla\lambda\nabla\cdot\vec{U}\\ +\nabla\mu\times\nabla\times\vec{U}+2(\nabla\mu\cdot\nabla)\vec{U}\label{eqn1}
\end{split}
\end{align}
here $\vec{U}(\vec{r})$ is the particle displacement vector, $\rho(\vec{r})$ is the local density, $\lambda(\vec{r})$ and $\mu(\vec{r})$ are the local Lam{\'{e}} constants, which are all functions of the spatial variables. To apply the $\vec{k}\cdot\vec{p}$ method to Eqns.(\ref{eqn1}), we write the Bloch functions in the vicinity of the $\Gamma$ point as linear combinations of the three degenerate states. After substituting these functions into Eqn.(\ref{eqn1}), applying periodic boundary conditions, and collecting only the linear order terms in $\vec{k}$ we obtain, 
\begin{equation}
det\bigg\vert (\omega_{n\vec{k}}^2-\omega_{j0}^2)I+P(\vec{k}) \bigg\vert=0
\label{eqn2}
\end{equation}
where $n$ denotes the band index at the Bloch wave vector $\vec{k}$, and P is the reduced Hamiltonian matrix with elements $P_{jl}=i\vec{k}\cdot \vec{p}_{jl}$, that represent the coupling strength between the degenerate state $j$ and $l$ at the $\Gamma$ point. The matrix element $\vec{p}_{jl}$ can be calculated as,
\begin{equation}
\begin{split}
p_{jl}(\vec{k})=\frac{(2\pi)^3}{V}\int_{unitcell}\Bigg \{ (\lambda+\mu)(\nabla \vec{U}_{j0})^{T}\cdot\vec{U}_{l0}^{*} +\\
(\lambda+\mu)[\nabla\cdot\vec{U}_{j0}(\vec{r})]\vec{U}_{l0}^{*}+2\mu\nabla\vec{U}_{j0}\cdot\vec{U}_{l0}^{*}+[\nabla\lambda\cdot\vec{U}_{l0}^{*}]\vec{U}_{j0}\\+2(\vec{U}_{j0}\cdot\vec{U}_{l0}^{*})\nabla\mu + [\nabla\mu\cdot\vec{U}_{j0}]\vec{U}_{l0}^{*}-[\vec{U}_{j0}\cdot\vec{U}_{l0}^{*}]\nabla\mu\Bigg \}
\end{split}
\label{eqn3}
\end{equation}
After evaluating Eqn.(\ref{eqn3}) numerically, the reduced Hamiltonian matrix is given by,

\begin{align} 
\vec{p}_{jl}=10^8\left(
\begin{smallmatrix} 
(0,0) & (0.002,-4.9983) & (-4.9902,-0.002)\\
(-0.002,4.9930) & (0,0) & (0,0)\\
(4.9849,0.002) & (0,0) & (0,0)
\end{smallmatrix}
\right) 
\label{eqn4}
\end{align}

\begin{figure}
	\includegraphics[scale=0.42]{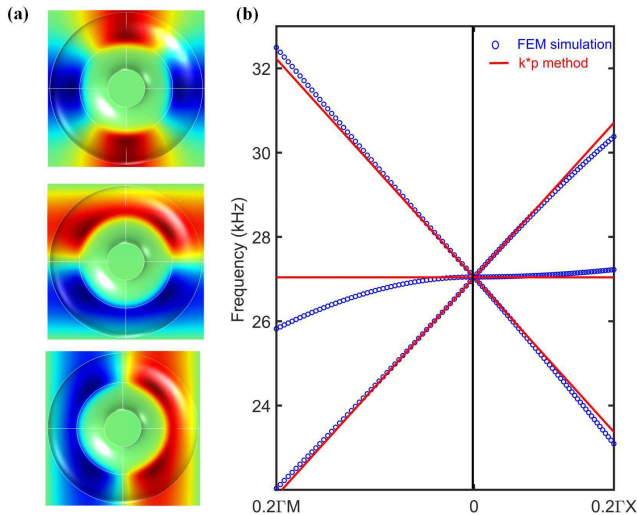}
	\caption{ (a) The eigenstates corresponding to the three degenerate states at the DP. From top to bottom, we find the state of the negative slope band, the flat band, and the positive band. (b) Comparison of the linear dispersion prediction at the DP obtained from the $\vec{k}\cdot\vec{p}$ method and the finite element simulations. }\label{Fig3}
\end{figure}

Note that $\vec{p}_{12}=-\vec{p}_{21}$, $\vec{p}_{13}=-\vec{p}_{31}$, $\vert p_{12}\vert=\vert p_{13}\vert$ and $\vec{p}_{12}\perp\vec{p}_{13}$; these properties are required to guarantee the isotropy of the cones. By substituting Eqn.(\ref{eqn4}) into Eqn.(\ref{eqn1}), we get the dispersion relations of the modes contributing to the Dirac-like cone,
\begin{equation} \label{eqn5}
\begin{split}
&\frac{\Delta f}{\Delta k}=0 \\ 
&\frac{\Delta f}{\Delta k}=\pm \frac{\vert\vec{p}_{12}\vert}{8\pi^2f_0}=\pm 233.48 
\end{split}
\end{equation}
where we have approximated the term $\omega_{n\vec{k}}^2-\omega_{j0}^2$ as $-2\omega_{j0}\Delta\omega$. Obviously the first result in Eqn.(\ref{eqn5}) corresponds to the flat band while the remaining two signed values correspond to the linear dispersion associated with the cones. Note that these results do not depend on the wave vector $\vec{k}$ direction thus confirming the isotropy of the linear dispersion. Equations (\ref{eqn5}) can be plot together with the numerically obtained dispersion relations (Fig.\ref{Fig3}b) in order to illustrate the good agreement between the $\vec{k}\cdot\vec{p}$ method and the full field numerical results. It can be seen from Eqn.(\ref{eqn4}) that the linear dispersion was determined by the non-diagonal element of the $p_{jl}$ matrix. This term represents the strength of the coupling between the degenerate states, and indicates that the frequency repulsion effect gives rise to the Dirac-like Cones.

\subsection{Effective medium properties}
An interesting feature of the Dirac-like cone is that, under certain conditions, it can be mapped to an effective medium. Reducing the dynamic properties to effective medium properties allows a very clear characterization of the double zero properties.
Note that although the selected DP has a relatively high frequency (the wavelength in the flat plate is about 1.25a), we employed the boundary effective medium theory \cite{Laihybridsolid} (see Supplementary material) to obtain the effective material parameters. It has been argued \cite{EM1,EM2} that, for periodic media, the effective medium theory \cite{EM3,EM4} is still valid at $\vec{k}=\vec{0}$ around the standing wave frequency even if the frequency belongs to the short wavelength regime. From a more empirical perspective, we will also show that the use of the effective medium description matches well the finite element model predictions. 

\begin{figure}
	\includegraphics[scale=0.42]{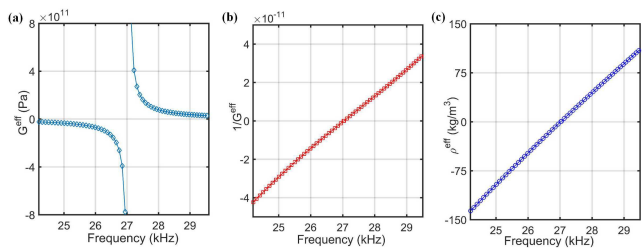}
	\caption{Frequency dependence of (a) the effective shear modulus $G^{eff}$, (b) the reciprocal effective shear modulus $1/G^{eff}$, and (c) the effective mass density $\rho^{eff}$ near the Dirac-like point.} \label{Fig4}
\end{figure}

By using the boundary effective medium theory \cite{Laihybridsolid}, we obtained the effective shear modulus and the effective mass density as function of frequency in the range around the Dirac-like point (Fig.\ref{Fig4}a and c). It can be seen that below the DP, the proposed metamaterial design acts as a double negative material while above the DP it acts as a double positive material. Similarly, the calculation of the reciprocal effective shear modulus (Fig.\ref{Fig4}b) shows that both $\rho^{eff}$ and $1/G^{eff}$ vary linearly and become zero simultaneously as the frequency crosses the Dirac-like point. These results suggest that the proposed metamaterial should exhibit propagation properties consistent with a double-zero-index material. In particular, we expect that acoustic waves traveling through the medium at the DP frequency should not experience any spatial phase change.

\begin{figure}
\includegraphics[scale=0.42]{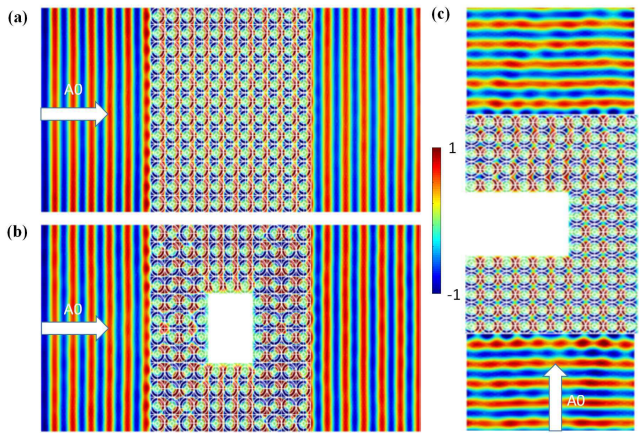}
\caption{ (a) The out-of-plane displacement distribution of the wave field when an incident $A_0$ planar wave at $f=27.04$ kHz impinges normally onto the $11\times12$ lattice of double-zero-index material. The top and bottom boundaries were treated with periodic boundary conditions while the left and right boundaries used PMLs to avoid reflections. As expected, no phase change occurs inside the metamaterial slab. (b) The out-of-plane displacement field showing the cloaking capability of the double zero material when an object $3\times4$ units in dimension is embedded in the slab. (c) The out-of-plane displacement field showing the supercoupling capabilities when a U-shaped narrow channel is filled with the double zero material.} \label{Fig5}
\end{figure}

\subsection{Full field numerical analyses}
In order to validate the theoretical predictions formulated above, we built a numerical model of an elastic waveguide made out of the proposed double-zero-index material. The basic test structure consisted in a flat plate with a $11\times12$ lattice of torus-like tapers embedded in the center (Fig.\ref{Fig5}a and b). Left and right boundaries were treated with perfectly matched layers (not shown) to avoid reflections, while top and bottom boundaries were treated with periodic conditions in order to simulate an infinite plate. The zero-index material slab was excited from the left by a planar $A_0$ wave at $f=27.04$ kHz and normal incidence. The resulting wave field (Fig.\ref{Fig5}a) indicated that no phase change occurred inside the metamaterial slab and nearly full transmission was achieved due to the zero-refractive-index and the matched impedance with the flat plate. These peculiar transmission properties were tested to show the ability to achieve cloaking and supercoupling in structural waveguides.

To illustrate the cloaking capability we embedded an object (represented by a through-hole opening $3\times4$ units in size) within the DZIM metamaterial. The opening had clamped boundary conditions. The remaining conditions were unchanged with respect to the case discussed above. The numerical results (Fig.\ref{Fig5}b) show that the wave emerges on the opposite side of the slab being completely unaffected. Comparing these results with Fig.\ref{Fig5}a, we see that the acoustic field downstream of the object does not carry any information about the object itself therefore confirming the cloaking capability of the medium. 

In a similar way, we tested the transmission performance of a U-shaped waveguide channel. The change in cross section was operated within the zero-index slab by shrinking the middle section from 10 to 4 units. Similarly to the cloaking case, clamped boundary conditions were imposed on all the walls of the U-shaped waveguide while the remaining boundaries were left free. In addition, perfectly matched layers were used on the top and bottom surfaces to absorb the outgoing waves and eliminate reflections. We observe that the incident plane wave ($A_0$ mode at $f=27.04$ kHz) propagated through the U-shaped channel completely unaffected picking up only a minor phase distortion. These results confirm the ability of the proposed design to create supercoupling effects in structural waveguides.

\subsection{Experimental Results}

We performed an experimental investigation in order to validate the concept of DZIM structural waveguide. We selected the supercoupling case for testing because it is the most challenging condition to achieve and therefore the most representative of the actual performance. To facilitate fabrication and testing, we rescaled the structure by reducing the plate thickness to $0.004$ mm and the unit cell dimension to half of the original size. The rescaling resulted in a Dirac-like point at $54.08$ kHz. The test sample was fabricated with extended edges along the U-Shaped waveguide channel in order to be able to enforce the boundary conditions. For simplicity, instead of creating a complex setup to impose the clamped conditions used in the simulation results (Fig.\ref{Fig5}c), we decided to treat these edges of the U-section with viscoelastic damping material while clamping the remaining edges. Numerical simulations were performed to show that this selection of boundary conditions ensured similar transport behavior to those used in the previous simulations (see supplementary material). 

\begin{figure}
	\includegraphics[scale=0.4]{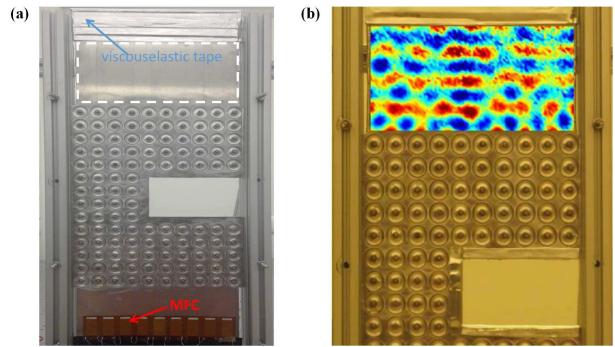}
	\caption{Experimental setup and results. (a) Front view of the testbed consisting of a $4mm$ thick aluminum plate with a U-shaped waveguide channel filled with the proposed DZIM material. An array of MFC patches was surface bonded to generate the ultrasonic excitation. The white dashed box indicates the area where the out-of-plane displacement response was measured. (b) The measured transmitted $A_0$ wave field (out-of-plane component) showing that the waves propagates through the U-shaped channel preserving its planar nature.} \label{Fig6}
\end{figure}

The torus-like tapers were CNC machined from an initially flat aluminum plate while the center masses were cut from aluminum bars and successively glued on the taper. The experimental sample was mounted vertically in an aluminum frame and viscoelastic tape was applied on the top and bottom edges in order to minimize reflections from the boundaries. An array of Micro Fiber Composites (MFC) patches (Fig.\ref{Fig6}a) was surface bonded on the plate and simultaneously actuated to generate a quasi $A_0$ planar incident wave. The excitation signal was a 50-count wave burst with a $54.1$ kHz center frequency. The out-of-plane response of the plate after the DZIM slab (marked by the dashed white box in Fig.\ref{Fig6}a) was acquired using a Polytec PSV-500 laser scanning vibrometer. Figure \ref{Fig6}b shows the measured out-of-plane velocity field at a given time instant. The measurements clearly indicate that the $A_0$ wave passes through the U-shaped waveguide channel preserving its planar wavefront. Note that the finite nature of the section before and after the metamaterial slab produces reflections that alter the structure of the transmitted wavefront. Despite the reverberation effect due to the finite boundaries, the planar nature of the transmitted wave fields is still well identifiable. These results confirm the theoretical and numerical predictions and the overall behavior of the double-zero-index waveguide.

\section{Conclusions}

We have presented and experimentally demonstrated the design of a double zero index structural waveguide. The design leverages the use of locally-resonant geometric tapers that are used as fundamental unit cells to achieve and tune Dirac-like dispersion at the center of the Brillouin zone. We showed, both by theoretical and numerical methods, that the material can be mapped to a double zero effective medium when excited in the neighborhood of the Dirac-like point. Full field numerical simulations showed that this material can be used to achieve cloaking and supercoupling in elastic waveguides. Both theoretical and numerical results were confirmed by experimental measurements that validated the design and provided conclusive evidence that double zero properties can be successfully achieved in solids.

\section{Acknowledgments}
The authors gratefully acknowledge the financial support of the Air Force Office of Scientific Research under the grant YIP FA9550-15-1-0133.

\bibliography{refbib}

\end{document}